\documentclass[twocolumn,amsmath,amssymb]{revtex4}
\usepackage[dvips]{graphicx}
\begin{document}
\newcommand{\gsim}{ \mathop{}_{\textstyle \sim}^{\textstyle >} }
\newcommand{\lsim}{ \mathop{}_{\textstyle \sim}^{\textstyle <} }
\newcommand{\vev}[1]{ \left\langle {#1} \right\rangle }
\newcommand{\bra}[1]{ \langle {#1} | }
\newcommand{\ket}[1]{ | {#1} \rangle }
\newcommand{\ev}{ {\rm eV} }
\newcommand{\kev}{{\rm keV}}
\newcommand{\mev}{{\rm MeV}}
\newcommand{\gev}{{\rm GeV}}
\newcommand{\tev}{{\rm TeV}}
\newcommand{\mpl}{$M_{Pl}$}
\newcommand{\mw}{$M_{W}$}
\newcommand{\Ft}{F_{T}}
\newcommand{\Zparity}{\mathbb{Z}_2}
\newcommand{\BLambda}{\boldsymbol{\lambda}}
\newcommand{\be}{\begin{eqnarray}}
\newcommand{\ee}{\end{eqnarray}}
	
   \title{Reduced Fine-Tuning in Supersymmetry with R-parity violation}
\author{Linda M Carpenter, David E Kaplan, Eun-Jung Rhee}
\affiliation{Department of Physics and Astronomy, 
		Johns Hopkins University, 
		Baltimore, MD  21218}
   \date{\today}                                                               
   \setcounter{footnote}{0}                                                    
   \setcounter{page}{1}                                                        
   \setcounter{section}{0}                                                     
   \setcounter{subsection}{0}                                                  
   \setcounter{subsubsection}{0}                                               
   \begin{abstract}
Both electroweak precision measurements and simple supersymmetric extensions of the standard model prefer a mass of the Higgs boson less than the experimental lower limit of 114 GeV.  We show that supersymmetric models with R parity violation and baryon number violation have a significant range of parameter space in which the Higgs dominantly decays to six jets.  These decays are much more weakly constrained by current LEP analyses and would allow for a Higgs mass near that of the $Z$.  In general, lighter scalar quark and other superpartner masses are allowed and the fine-tuning typically required to generate the measured scale of electroweak symmetry breaking is ameliorated.  The Higgs would potentially be discovered at hadron colliders via the appearance of new displaced vertices.  The lightest neutralino could be discovered by a scan of vertex-less events LEP I data.
   \end{abstract}
   \maketitle                                                                  
The Standard Model of particle physics is arguably the crowning achievement of the last half-century's work towards the understanding of the laws of nature at short distances.  However, two somewhat nagging features remain.   The first is that while statistical fits of standard model parameters to precision measurements produce a best fit value for the Higgs scalar mass of $85^{+39}_{-28}$ GeV \cite{LEPEWWG}, LEP II places a lower bound of 114.4 GeV at 95\% CL \cite{unknown:2006cr}. While these constraints taken together do not constitute a discrepancy, a Higgs mass measured below the current LEP bound would have improved the fit to precision data.  In addition, it has been argued that the electroweak observables most sensitive to the Higgs mass are themselves not in good agreement and imply a discrepancy with the LEP II bound \cite{Chanowitz:2002cd}.  The second feature is that the scale of electroweak symmetry breaking is very sensitive to quantum corrections.  If the standard model is valid to some very high energy scale $M \gg 1$ TeV, the parameters of the ultraviolet theory would require an unnatural tuning of order one part in $(M/ 1{\rm\; TeV})^2$ to maintain the hierarchy.  While this fact alone does not guarantee new physics beyond a Higgs boson at the electroweak scale, it is strongly suggestive of physics at the weak scale which stabilizes scalar masses with respect to radiative corrections.

A well-known solution to the naturalness problem is to impose supersymmetry on the standard model and softly break it at a scale of $M\sim 1$ TeV (for a review, see \cite{Martin:1997ns}).  Radiative corrections to scalar masses in these theories are proportional to the scale of supersymmetry breaking and therefore naturally stabilize the mass of the Higgs at around the weak scale.  Remarkably, the minimal version of these theories predict the unification of couplings at a renormalization scale near the Planck scale.  A discrete symmetry, R parity, is introduced to forbid dimension-four baryon and lepton number violating operators and avoid proton decay as we discuss below.  

While the MSSM (Minimal Supersymmetric Standard Model) contains over a hundred new parameters, it has become tightly constrained.  A robust constraint on the MSSM is the bound on the Higgs mass.  The physical mass gets contributions which depend only logarithmically on superpartner masses (for example, the scalar top quark mass) through corrections to the Higgs quartic interaction.  On the other hand, the $Z$ boson mass and the scale of electroweak symmetry breaking gets corrections {\it proportional} to superpartner masses.  To satisfy the current bound on the physical mass, large scalar top masses ($m_{\tilde t}\simeq 1$ TeV) are required.  For a large cutoff $\Lambda$, say of order the Planck scale, contributions to the (squared) $Z$ mass will be roughly of order the superpartner masses, say $\delta m_Z^2\sim m_{\tilde t}^2$.  A cancelation would be required among contributions with a tuning of the order one part in $(m_{\tilde t}/m_Z)^2$.  Thus, 1 TeV scalar tops would require $\sim 1\%$ tuning.  A beautiful discussion of this tension in the MSSM is contained in \cite{Giudice:2006sn}.

One possible resolution to the paradox is that the Higgs is in fact light but missed by experiments.  The quoted lower bound on the Higgs mass comes from analyses assuming a Higgs with standard model properties such as a standard model cross section for $Z$-Higgs production and standard model branching ratios into bottom quarks and tau leptons.  If the branching ratio to standard model final states are uniformly suppressed by, for example, a factor of five -- and the new decay modes are not picked up by any LEP searches -- the 95\% CL lower limit on the Higgs mass reduces to roughly 93-95 GeV (see Figure 2 of \cite{unknown:2006cr}).  Our model exploits this weakness.  Other attempts to modify Higgs decays for the purpose of naturalness have been made in the context of the next-to-minimal supersymmetric standard model \cite{Dermisek:2005ar}, and in a general analysis of the MSSM with an additional singlet superfield \cite{Chang:2005ht}.

In this letter, we show that in the MSSM with R parity violation and non-unified gaugino masses, there is a significant amount of parameter space in which the Higgs dominantly decays to a pair of unstable neutralinos, each of which subsequently decays to three quark jets. The parameter space allows, as we detail below, Higgs masses around the $Z$ mass even with a standard model production cross section;  this is our main result.

R parity is a symmetry under which all superpartners are odd.  It forbids the following renormalizable operators in the superpotential:
\begin{eqnarray}
 W & \supset &
 \mu_i L_i {\bar H} + \lambda_{ijk} L_i L_j E^c_k + \lambda'_{ijk} L_i Q_j D^c_k \nonumber \\ & + & \lambda''_{ijk} U^c_i D^c_j D^c_k ,
 \label{eq:rpv}
 \end{eqnarray}
 where $L$, $E^c$, $\bar{H}$, $Q$, $U^c$, and $D^c$ are lepton doublet, lepton singlet, up-type Higgs, quark doublet, up-type quark singlet, and down-type quark singlet superfields respectively, and the $ijk$ are flavor indices.  These interactions violate lepton number (the first line) and baryon number (the second).  Their existence would predict unacceptable levels of proton decay unless at least some of these couplings are extremely small.  However, proton stability could be provided by a symmetry that allows only the lepton-number or baryon-number violating terms \cite{Ibanez:1991pr}.  In this paper, we focus on the latter.  Bounds on the individual $\lambda''$ couplings are only stringent from neutron--anti-neutron oscillations and double nucleon decay, requiring $\lambda''_{112}\lsim 10^{-7}$ and $\lambda''_{113}\lsim 10^{-4}$ for 200 GeV scalar quark and gluino masses.  The other seven coupling are less constrained.  The tightest bounds are on products of two different couplings which range from $\lambda''_{ijk}\lambda''_{i'j'k'}<10^{-2} -10^{-4}$ and come dominantly from limits on rare hadronic decays of $B$ mesons.  For a broad review of R parity violation in supersymmetry, see \cite{Barbier:2004ez}.

Do LEP searches put a bound on a Higgs that decays to 6 quarks (via two neutralinos)?  No analysis has been performed looking for this exclusive final state.  A decay-mode-independent search for a Higgs boson was performed by the OPAL experiment (by looking for the associated $Z$ in leptonic channels \cite{Abbiendi:2002qp}) and puts a lower bound of 82 GeV when the production cross section is equal to that of the standard model.  In addition, the search for a Higgs decaying to two jets of any flavor \cite{:2001yb} could be sensitive to our Higgs to six jets when the latter can be forced into a two-jet topology (there may also be sensitivity from $h\rightarrow 2b$ when each neutralino decay contains a $b$ quark).  An analysis of this type was done by DELPHI \cite{Abdallah:2004wy} in the search for a cascade decay of the Higgs to four $b$ quarks via two pseudo scalars, $a$ \cite{opal-note}.  They modified the search for $e^+e^-\rightarrow hZ\rightarrow (b\bar{b})Z$ to be sensitive to the cascade $h\rightarrow aa\rightarrow b\bar{b}b\bar{b}$ by forcing the latter into two jets and estimating the efficiency of the $h\rightarrow 2b$ search to pick up $h\rightarrow 4b$.  Comparing Tables 27 and 29 in \cite{Abdallah:2004wy}, one can see that DELPHI rules out a 100 GeV Higgs decaying exclusively to $b\bar{b}$ with roughly a quarter of the standard model cross section, while it requires 80\% of the cross section to rule out the same mass Higgs decaying to $4b$'s (about three times the signal events).  In our case, the Higgs decays to six jets and thus the jets will be softer and more numerous and thus harder to reconstruct into two jets.  To see the effect of softer jets, the same tables show that for a 65 GeV Higgs decaying to $4b$'s vs. $2b$'s requires 7-8 times more signal events -- and this is even with pseudo scalars as light as 12 GeV.  In addition, part of the DELPHI analysis takes advantage of the additional $b$ quarks in the final state, whereas our final state will have at most $2b$'s.  Assuming the loss of the extra $b$ quarks in the final state cuts the efficiency in half, we estimate that our soft jets with 2 $b$'s in the final state would be picked up by the $2b$ search with an efficiency of $\sim 1/7 \times 1/2 \sim 7\%$.  This would also be true of the flavorless search \cite{:2001yb}, which is similar to the $2b$ search with the $b$-tagging requirement removed.  While we feel this estimate is conservative, it is very rough and a full analysis is warranted.

What are the constraints on the mass of the lightest neutralino?  The neutralino should be light enough to allow for our Higgs decay ($\lsim 50$ GeV), while the lightest chargino must satisfy its current lower bound ($\sim 103$ GeV in most of parameter space, even for R-parity violating decays \cite{Achard:2001ek}).  This constrains the MSSM parameters such that $M_1< M_2,\mu$, and the lightest neutralino is mostly bino -- although it must have enough of a higgsino component to allow the Higgs width to be dominated by this decay, and thus the $\mu$ parameter shouldn't be too much larger than 100 GeV as we see below (see also \cite{Cavalli:2002vs}).  

The remaining question then is how could such a light neutralino with strong enough couplings to dominate the Higgs width not being detected indirectly by its effect on the $Z$ width or directly in searches at LEP II.  There are two reasons:  the first is that the width of the $Z$ in the standard model ($\sim 2.5$ GeV) is three orders of magnitude bigger than the standard model width of a 100 GeV Higgs.  The second is that in the range of small to moderate bino-higgsino mixing, the Higgs decay rate into neutralinos is roughly proportional to the mixing angle squared while the same rate for the $Z$ goes like the mixing angle to the fourth power.  The decay width of the $Z$ into the lightest neutralino at tree level is
\begin{equation}
\Gamma_{\chi^0} = \Gamma_{\nu}\times \Delta^2  \sqrt{1 - \left(\frac{2 m_\chi}{m_Z}\right)^2}\left( 1-\left(\frac{m_\chi}{m_Z}\right)^2 \right)
\end{equation}
where $\Delta$, defined in the appendix, is roughly the bino-higgsino mixing angle squared, and $\Gamma_\nu$ is the standard model $Z$ width into one species of neutrino.  We require this contribution to the total and hadronic widths to be less than $0.1\%$ -- roughly $1\sigma$ as determined by the electroweak fit \cite{LEPEWWG}.  This requirement sets a bound of $\Delta\lsim 1/10$ for a very light neutralino, and weaker for heavier neutrinos as phase space gets reduced.  We find that in most of our parameter space -- where the decay to neutralinos dominates the Higgs width and the chargino bound is satisfied -- we satisfy this constraint.

Searches for neutralinos which decay via baryon number violation have been preformed by Aleph, DELPHI and L3 \cite{Achard:2001ek}.  None of the three searches were able to put a bound on the neutralino mass via a direct search, but only through a search for a chargino and the theoretical assumption $M_1=(5/3)\tan^2{\theta_W} M_2\sim M_2/2$ relating the two masses through the assumption of gaugino mass unification.  The L3 experiment does present cross-section bounds for neutralino masses between 30 GeV and roughly 100 GeV of around 0.1 pb.  The neutralino cross section through an s-channel $Z$ at LEP II is
\begin{equation}
\sigma_{\rightarrow \chi^0\chi^0} = \sigma_{\rightarrow\nu\bar{\nu}} \times \Delta^2 \sqrt{1 - \frac{4 m_\chi^2}{s}}\left( 1-\frac{m_\chi^2}{s} \right) ,
\end{equation}
where $\sigma_{\rightarrow\nu\bar{\nu}}$, the neutrino pair-production cross section, is $\sim 1$ pb at center of mass energy $\sqrt{s}=200$ GeV.  To satisfy the L3 bound, we require $\Delta<1/3$.  This bound is satisfied in our entire parameter space.  However, if scalar leptons are relatively light, a t-channel diagram can dominate the cross section and overwhelm the bound.  Requiring the cross section to satisfy the L3 constraint places a lower bound of $\sim 300$ GeV on scalar electron masses (in the case degenerate scalars).  This becomes our strongest constraint on a superpartner mass in the baryon-number violating MSSM.  The lightest neutralino, due to the weakness of their couplings and the nature of their decays, has no significant collider bound within our parameter space.

\begin{figure}
\centerline {\includegraphics[width=2.7 in]{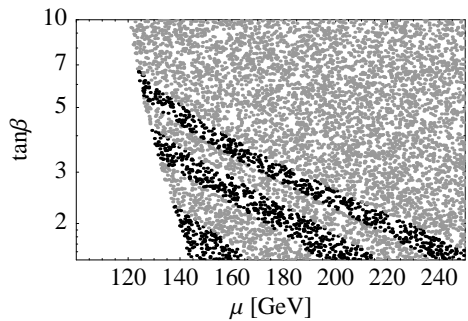}}
\caption{A random scan of the parameters $\mu$ and $\tan\beta$ with $M_2=250$ GeV, $M_1=50$ GeV and $m_{higgs} = 100$ GeV.  The borders between the alternating black and gray points represent Higgs branching ratios to neutralinos of (from left to right) 90\%, 85\%, 80\%, 75\%, and 70\% respectively.  The white space to the left is excluded by the chargino mass bound.}
\label{fig:mu-tanb}
\end{figure}

\begin{figure}
\centerline {\includegraphics[width=2.6 in]{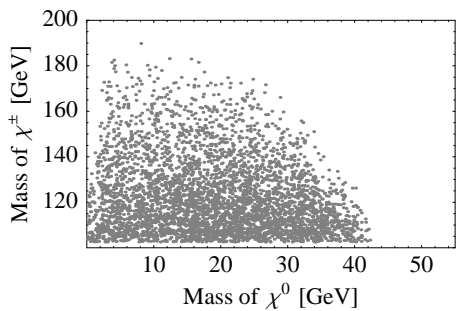}}
\caption{The lightest chargino mass versus lightest neutralino mass in a scan of $\mu$ (from 120 to 250 GeV), $\tan\beta$ (2 to 5), $M_1$ (10-100 GeV), and $M_2$ (150-400 GeV).  For all points, the branching ratio to neutralinos is at least 75\%.}
\label{fig:mx-mch}
\end{figure}

The points in parameter space which predict a large Higgs to neutralinos branching ratio and satisfy the lower bound on the chargino mass satisfy $M_2 > 3 M_1$ \cite{Cavalli:2002vs}, and the effect on the branching ratio becomes unimportant above $M_2>250$ GeV.   In Figure \ref{fig:mu-tanb} we show a plot of different branching ratios of the Higgs to neutralinos for $M_2$ , $M_1$, and the Higgs mass fixed at 250, 50, and 100 GeV respectively.  Each point also satisfies the constraint on the contribution to the hadronic $Z$ width.  We also require $m_{\chi^0}>12$ GeV to allow for significant phase space for the decay.  We see that a large branching ratio requires relatively low values of $\tan\beta$ and $\mu$.  In Figure \ref{fig:mx-mch} we scan over $M_1$, $M_2$, $\mu$, and $\tan\beta$ and plot points which satisfy the chargino mass and $Z$ width bounds.  For these points, the branching ratio is less than 25\% to normal standard model decays, thus lowering the Higgs mass bound in this part of parameter space to roughly 95-100 GeV according to Figure 2 of \cite{unknown:2006cr}.

The scans are done in the decoupling limit ({\it i.e.}, the pseudo-scalar mass is fixed at 1 TeV), where the heavier CP-even Higgs boson is much more massive and thus all couplings of this lightest Higgs are standard model like.  Away from this limit, the decay width to standard model channels increases while the overall production cross section goes down.  Moderate mixing with the heavier Higgs does not significantly change the qualitative features of these plots.

The decay length of the neutralino can be long enough to leave a displaced vertex.  The average decay length of the lightest neutralino is \cite{Dreiner:1991pe}
\begin{eqnarray}
L & \simeq & \frac{384\pi^2 \cos^2\theta_w}{\alpha \left| U_{21}\right|^2 \lambda''^2}\frac{m_{\tilde{q}}^4}{m_\chi^5} ( \beta\gamma)\\\nonumber
& \sim & \frac{3 {\mu}m}{\left| U_{21}\right|^2} \left(\frac{10^{-2}}{\lambda''}\right)^2 \left(\frac{m_{\tilde q}}{100\; {\rm GeV}}\right)^4 \left(\frac{30\; {\rm GeV}}{m_\chi}\right)^5 \frac{p_\chi}{m_\chi},
\end{eqnarray}
where $|U_{21}|$ is an element of the mixing matrix in the appendix and $p_\chi$ is the neutralino's momentum.  Final-state particle masses, Yukawa couplings and QCD corrections have all been neglected.  For very small couplings, light neutralinos, or heavy scalar quarks, the decay length could be quite long and might have been seen as anomalous events at LEP if they decay in the tracking chamber, and perhaps by searches for stable squarks and gluinos \cite{Heister:2003hc} if they decay in or near the hadronic calorimeter.  If their decay length is longer than about a meter, the invisible Higgs search would pick up these events and rule out masses up to 114 GeV \cite{unknown:2001xz}.

What does baryon number violation do for susy parameter space?  In general, bounds on superpartner masses are weaker than in the case of the R-parity conserving MSSM due to a lack of missing energy in the signal.  The current bounds of 200 - 300 GeV on squarks and gluinos from Tevatron searches came from analyses which required significant missing transverse energy cuts \cite{Abazov:2006bj}.  With baryon-number and R-parity violating interactions, the bounds on all superpartners are below 100 GeV, except for the chargino, whose bound remains roughly the same (102.5 GeV) \cite{Eidelman:2004wy}.  The bounds on the lightest neutralino quoted in the particle data book are due to {\it chargino} searches and the requirement of gaugino mass unification.  The direct search at LEP for a decaying lightest neutralino is unable to put bounds on its mass.  

Of course another impact of this model is the allowance of a lighter Higgs mass thus reducing the need for large radiative corrections to the quartic potential from the stop loop.  For the same value of $\tan\beta = 3$, the allowed lighter Higgs mass (say around 96 GeV) requires an enhancement of the quartic of only half as much as in the MSSM with R-parity conservation.  If instead we compare allowed MSSM Higgs masses at large $\tan\beta$ to our model's allowed Higgs masses at $\tan\beta=3$ (since we require low values for our decay to dominate), we still typically require a lower quartic enhancement by roughly $10-30\%$.  This translates into lower stop masses needed and less tuning.  However, while
R parity violation and non-unified gaugino masses help to relieve much of the persistent fine tuning in the MSSM, they clearly do not eliminate it \cite{Schuster:2005py}.  Among the strongest constraints are the chargino mass bound and the restrictions on contributions to $b \rightarrow s\gamma$.  In addition, avoiding the Higgs mass bound requires one to be in a non-generic part of parameter space in which the Higgs decays to neutralinos.  

R parity violation can allow for other non-standard Higgs decays which evade LEP searches.  For example, one linear combination of scalar bottom squarks can be perhaps as light as 7.5 GeV \cite{Janot:2004cy} due to suppressed couplings to the $Z$.  With baryon number violation, and sbottom masses below half the Higgs mass, this would allow the Higgs to decay to four light jets, and the decay would dominate standard Higgs decays at moderate to large $\tan\beta$ \cite{Carena:2000ka}.  On the other hand, lepton number violation through, for example,  the superpotential operator $\lambda'_{i33} L_i Q_3 D^c_3 $ could produce a dominant Higgs decay of $h\rightarrow 4b\; + \not{\!\!\! E}$ to which the standard $2b$ and $4b$ searches should have significantly reduced sensitivity.  These and other lepton-number violating decays are being explored \cite{wip}.

If the above scenario is correct, searching for the Higgs at hadron colliders could pose great difficulty.  However, if the neutralinos decay at a displaced vertex with a decay length greater than about 50 microns, these events could potentially be picked up by a dedicated search at the Tevatron, LHC, or LHCb \cite{petar,Strassler:2006ri}.  The vertex tagging at LHCb would be well suited for this search, and the statistics high enough - roughly 30\% of the Higgs bosons produced via gluon fusion are expected to fall in the detector's acceptance range \cite{Currat:2001mb}.  In addition, half of these decays would be baryon violating (assuming the lightest neutralino is a Majorana particle) and this could potentially be a striking signal.  Finally, the small but non-zero coupling of long-lived neutralinos to the $Z$ may allow them to be discovered by studying the beam gas (``vertex-less") events in LEP I data \cite{morris}.

{\bf Appendix:}
The neutralino mass matrix
\begin{align}
&U 
\left(\begin{array}{cccc}m_{\chi_1} & 0 & 0 & 0 \\0 & m_{\chi_2} & 0 & 0 \\0 & 0 & m_{\chi_3} & 0 \\0 & 0 & 0 & m_{\chi_4}\end{array}\right) U^T
= \\\nonumber 
&\left(
\begin{array}{cccc}
M_2 & 0 & m_Z c_w c_\beta & -m_Z c_w s_\beta \\0 & M_1 & -m_Z s_w c_\beta & m_Z s_w s_\beta \\m_Z c_w c_\beta & -m_Z s_w c_\beta & 0 & -\mu \\-m_Z c_w s_\beta & m_Z s_w s_\beta & -\mu & 0
\end{array}
\right) 
\end{align}
is diagonalized from the gauge basis to the mass basis by the orthogonal matrix $U$.  The eigenvalues are in ascending order in magnitude, and thus $m_{\chi_1}$ is the mass of the lightest neutralino.  In the gauge basis, the mass matrix above multiplies the vector $\{\tilde{W},\tilde{B},\tilde{H},\tilde{\bar{H}}\}$ corresponding to the wino, bino and down- and up-type higgsinos.  The bino-higgsino mixing can be characterized by a parameter $\Delta$ (used in the text) defined as
\begin{equation}
\Delta=\left| U_{13}\right|^2 - \left| U_{14}\right|^2
\end{equation}

\vskip 0.15 in
{\bf Acknowledgments} 
\vskip 0.15 in
We thank P. Maksimovic, M. Swartz and N. Weiner for useful conversations,Y. Gao for collaboration in early stages of this work, and R. Sundrum for reading the draft.  This work  is supported in part by NSF grants PHY-0244990 and PHY-0401513, by DOE grant DE-FG02-03ER41271, and by the Alfred P. Sloan Foundation.


\bibliographystyle{apsrev}

\end{document}